\documentclass[onecolumn]{mn2e}
\usepackage{graphicx}

\def\Msun{\>{\rm M_{\odot}}}

\newcommand{\gtsim}{\mathrel{\hbox{\rlap{\lower.55ex \hbox {$\sim$}}
                   \kern-.3em \raise.4ex \hbox{$>$}}}}
\newcommand{\ltsim}{\mathrel{\hbox{\rlap{\lower.55ex \hbox {$\sim$}}
                   \kern-.3em \raise.4ex \hbox{$<$}}}}

\title{Supermassive Black Hole Growth and Merger Rates from Cosmological N-body Simulations}

\author[Miroslav Micic, Kelly Holley-Bockelmann, Steinn Sigurdsson, Tom Abel]
{Miroslav Micic$^1$\thanks{E-mail: micic@astro.psu.edu, kellyhb@gravity.psu.edu,
steinn@astro.psu.edu, tabel@slac.stanford.edu}, Kelly Holley-Bockelmann$^2$, Steinn Sigurdsson$^1$ \& Tom
Abel$^3$ \\
$^1$ Department of Astronomy \& Astrophysics, Pennsylvania State University  \\
$^2$ IGPG, Pennsylvania State University  \\
$^3$ SLAC, Stanford University \\
}

\begin{document}
\maketitle

\begin{abstract}

Understanding how seed black holes grow into intermediate and supermassive
black holes (IMBHs and SMBHs, respectively) has important implications for
the duty-cycle of active galactic nuclei (AGN), galaxy evolution, and
gravitational wave astronomy. Most studies of the cosmological growth and
merger history of black holes have used semianalytic models and have
concentrated on SMBH growth in luminous galaxies. Using high resolution
cosmological N-body simulations, we track the assembly of black holes over
a large range of final masses -- from seed black holes to SMBHs -- over
widely varying dynamical histories. We used the dynamics of dark matter
halos to track the evolution of seed black holes in three different gas
accretion scenarios. We have found that growth of a Sagittarius A* - size
SMBH reaches its maximum mass M$_{\rm SMBH}$=$\sim$10$^6\Msun$ at z$\sim$6
through early gaseous accretion episodes, after which it stays at near
constant mass. At the same redshift, the duty-cycle of the host AGN ends,
hence redshift z=6 marks the transition from an AGN to a starburst galaxy which
eventually becomes the Milky Way. By tracking black hole growth as a function
of time and mass, we estimate that the IMBH merger rate reaches a maximum of
R$_{\rm max}$=55 yr$^{\rm -1}$ at z=11. From IMBH merger rates we calculate
N$_{\rm ULX}$=7 per Milky Way type galaxy per redshift in redshift range
2$\ltsim$z$\ltsim$6.

\end{abstract}

\begin{keywords}

stars: Population III, intermediate mass black holes, supermassive black holes, 
gravitational waves, dark matter halos

\end{keywords}

\section{INTRODUCTION}

Supermassive black holes (SMBH) are thought to dwell at the centers of most
galaxies, (Kormendy $\&$ Richstone 1995) with masses between
10$^6 \, \Msun \ltsim$ M $\ltsim 10^9 \, \Msun$. In principle, the abundance
of SMBHs today can be explained if they grow through mergers and early
accretion (Schneider et al. 2002), from a gaseous disk. The most likely
candidates for SMBH seeds are black holes that form as remnants of Population
III stars at redshifts z$\gtsim$12-20 (Heger et al. 2003, Volonteri et al. 2003,
Islam et al. 2003, Wise $\&$ Abel 2005). These relic seeds are predicted to
form in the centers of dark matter halos (DMH), and have masses
$\ltsim 10^3 \Msun$ (Abel et al. 2000, 2002). DMHs form in the early universe
and hierarchically merge into larger bound objects. As DMHs merge into
massive halos, the seed black holes sink to the center through dynamical
friction and eventually coalesce.

Although the seed formation stops at z$\sim$12  as Population III
supernovae rates drop to zero (Wise $\&$ Abel 2005), SMBH growth continues
as DMH mergers proceed to low redshifts. From a combination of gas accretion
and binary black hole coalescence, seeds can grow to intermediate mass black
holes (IMBHs, with masses $10^2\,\Msun \ltsim$ m $\ltsim 10^5 \,\Msun$).
With continued mergers and gas accretion, it is thought that these IMBHs may
form the SMBHs we observe today. The detection of IMBHs is a matter of debate,
but possible candidates are ultraluminous X-ray sources in young star-forming
regions (Fabbiano 1989, Roberts $\&$ Warwick 2000, Ptak $\&$ Colbert 2004,
Fabbiano $\&$ White 2006) and nearby extragalactic star clusters.

Although this scenario works in general, real detailed understanding
of how seed black holes grow is a remaining challenge. For example,
there are still debates on the seed black hole mass, on the mass of
halos that can host seeds, on the seed formation redshift, on the
type and efficiency of gas accretion, and on the dynamics of the black
hole mergers, all of which compound to yield a huge range in
black hole merger rate estimates. Recent black hole merger rate calculations
span over three orders of magnitude: Haehnelt (1994) calculates the
merger rate to be R$\sim$ 0.1 - 100 events per year; Menou et al. (2001),
R$\sim$ 1 - 100 yr$^{\rm -1}$; Wyithe $\&$ Loeb (2003), R$\sim$ 15 -
350 yr$^{\rm -1}$; Sesana et al. (2004), R$\sim$ 10 yr$^{\rm -1}$; and
Rhook $\&$ Wyithe (2005), R$\sim$ 15 yr$^{\rm -1}$.

There are two general approaches to the problem: direct cosmological
N-body simulations and analytical techniques based on a Press-Schechter
(PS) formalism (Press $\&$ Schechter 1974). Both methods extract merger
rates from the DMHs' merger tree. N-body simulations have the advantage
that the evolution of density fluctuations is followed in complete
generality, without the need for any of the assumptions involved in
creating PS merger trees. Extended Press-Schechter theory (EPS) combines
the PS halo mass function with halo merger rates derived by
Lacey $\&$ Cole (1993), and it stands as the most widely used method for
calculating merger rates. Unfortunately, EPS is mathematically inconsistent
(Erickcek et al. 2006) since it provides two equally valid merger rates
for the same pair of DMHs. It is also unclear whether EPS reproduces the DMH
mass function at high redshifts (Reed et al. 2007). With N-body simulations,
there are, in principle,  no constraints on the halo mass, structure or
kinematics. Unfortunately, the mass and spatial resolution needs to be
extremely high, and even in the best resolved N-body simulations of a large
volume $\Lambda$CDM universe (Nagashima et al. 2005), the mass resolution
sets the minimum DMH mass to 3$\times$10$^9$$\Msun$ for halos with as few
as ten particles. Hence using direct N-body simulations to track the
dynamics of seed black holes as they grow within a large comoving volume is
out of reach of our current technology.

In this paper, we have developed a hybrid method to follow the merger
history of seed black holes as they grow. We performed a high resolution
cosmological N-body simulation in the unexplored parameter space of a
{\it small cosmological volume} but with very high mass resolution to
achieve well-resolved halos with a minimum of 32 particles and mass as
low as M$_{\rm halo}$=2.8$\times10^7\Msun$. The goal was to look at a
representative ``Local Group'' structure, comparable to the one which
hosts the Milky Way, and to resolve as low a mass as feasible.
We improved the algorithms for identifying dark matter structures and
developed set of physically-motivated criteria for seeding DMHs.
We then constructed the DMH merger tree, which provides a testbed to
study the effects of different gas accretion scenarios. In addition
to the merger rate, we extracted observables such as the black hole mass
function over cosmological time. This method is a major improvement in
calculations of MBH merger rates and presents the first step toward a full
treatment of the black hole growth problem. Massive black holes mergers will
be one of the prime signals for future space-based gravitational wave observatories like LISA
(Laser Interferometer Space Antenna) and BBO (Big Bang Observer).
Possible detection depends on a number of parameters: the mass ratio of
merging black holes, the total mass of the black hole binary and the redshift.

We describe our simulation and DMH seeding criteria in Section 2;
obtaining DMH and black hole merger trees in Section 3 and 4 together
with description of black hole growth models; black hole merger rates
in Section 5; finally, we discuss our results in Section 6.

\section{SIMULATION}

\subsection{Simulation Setup}

In our numerical simulations, we use GADGET (Springel et al. 2001)
to evolve a comoving 10 Mpc$^3 $ section of a $\Lambda$CDM universe
($\Omega_{\rm M}$=0.3, $\Omega_{\Lambda}$=0.7 and h=0.7) from $z=40$
to $ z=0$. We refine a sphere of 2 Mpc in the box to simulate at a
higher resolution with 4.9$\times10^6$ high-resolution particles
(softening length 2 kpc comoving). The rest of the box has
2.0$\times10^6$ low-resolution particles (softening length 4 kpc
comoving). The mass of each high resolution particle in this simulation is
8.85$\times10^5$M$_\odot$, and the mass of each low-resolution particle is
5.66$\times10^7$M$_\odot$. A detailed description of this simulation can be
found in Micic et al. 2006.

\subsection{Linking Length Problem}

Identifying and following a real DMH in a cosmological simulation
is technically challenging. If the minimum number of particles in a
halo is set too low, the DMH identified at one time-step might
later disperse. In the widely used Friends-of-Friends algorithm (FOF),
if the linking length (b) is too large, large dark matter structures
might bridge through a small number of tidally stripped particles.
When this occurs, the FOF algorithm can not properly identify DMHs.
A linking length of 0.2 has traditionally been used (Davis et al. 1985)
since the DMH mass function obtained with it proved to be consistent
with observations at low redshifts and mass functions obtained by PS
theory. However, very little is known about DMH mass functions at high
redshifts z$\gtsim$10 and low masses M$\ltsim$10$^{9}$$\Msun$, or about
the validity of this linking length. For example, Gao et al. (2005)
noted that the first massive halos are found in regions containing many
smaller halos which line up along filaments and sheets and can not be
identified by the FOF algorithm when a linking length of b=0.2 is used.
The only criterion for initializing linking length comes from the spatial
resolution in the performed numerical simulation. The linking length has
to be larger than the softening length. High resolution simulations have
small softening lengths, which allow the linking length to be smaller
as well. Consequently, in our high resolution simulation, we set b=0.1.
With this value for linking length we eliminated all the problems in
identification of dark matter structures.

\subsection{Analysis Tools}

We used P-GroupFinder algorithm (Springel 2000) to identify DMHs in
each of our 104 GADGET output files. The analysis starts with a FOF
algorithm (b=0.1) which spatially separates dark matter halos. The
minimum number of particles per halo is set to 32. The minimum DMH mass
resolved with 32 particles in our simulation is
M$_{\rm halo}$=2.8$\times10^7\Msun$. Then, we find a gravitationally
bound counterpart for each FOF halo and continue with the SUBFIND
algorithm to identify substructures in every DMH. We require that
halos and subhalos be virialized to be considered real. This
requirement is met when the DMH density is above 200 times the average
density of the universe at a specific redshift. Since we are studying
merger rates for Pop III massive black holes, we then select those
halos that are capable of hosting Pop III stars, using primordial
supernova rates (Wise $\&$ Abel 2005). Primordial stars can form in
DMHs with a minimum mass of 4.16$\times$10$^6$ $\Msun$
if H$_{\rm 2}$ cooling is the primary mechanism and if the UV
background has a negative feedback. They can also form through
hydrogen atomic line cooling in halos with mass
M$_{\rm vir}$$\gtsim$10$^8$$\Msun$[(1+z)/10]$^{-1.5}$. The lifetime
of Pop III stars is $\sim$3 Myrs and the end product is a population
of seed black holes with 100$\Msun$$\ltsim$M$_{\rm BH}$$\ltsim$1000$\Msun$.
The first Pop III seed black holes form at redshift z$\sim$20 and the
last Pop III stars die at redshift z$\sim$12. We select the interval
12$\ltsim$z$\ltsim$19 to seed those DMHs which have masses
M$_{\rm vir}$$\gtsim$10$^8$$\Msun$[(1+z)/10]$^{-1.5}$. At z=19 all
DMHs with mass above M$_{\rm halo}$=3.5$\times10^7\Msun$ and at z=12
all of the DMHs with mass above M$_{\rm halo}$=6.7$\times10^7\Msun$
are seeded.

\section{Dark Matter Halos' Merger Tree}

Starting with the first snapshot `` i'' , we identify DMHs with
the set of criteria described above. The most bound particle at
the center of each DMH that satisfies the criteria is set to be
a seed black hole. At the following snapshot `` i+1'' , we identify
halos from the previous ``i'' snapshot. At ``i+1'' we again seed new
halos that satisfy our criteria. By ``new'' halos we consider
only those halos that do not have any particles that were part
of any already seeded halo. This is because once a halo is seeded
it can not form a new black hole due to feedback from the UV
background of Pop III stars (Machacek et al. 2001). Meanwhile,
we also trace the parent DMH at ``i'' for every particle in
every DMH at ``i+1'', thus constructing a complete merger history.
The method is repeated through the redshift range 12$\ltsim$z$\ltsim$19.
Seeding stops at redshift z=12 but we continue with creating a
merger history until redshift z=0. In total we have 1447 seed black holes.

\section{Assembling a Black Hole}

After constructing the DMH merger tree, we turn our attention to
the evolution of the seeds. The N-body simulation tracks all the
seed black holes and their host halos from z=19-0. In order to
study black hole growth as a function of time, environment, and
merger type, we constructed three black hole merger models which
we detail below.

We parameterize the degree of gas accretion involved in growing
the black hole and the efficiency of binary coalescence. In the
fiducial case, we modeled black hole mergers in the simplest
possible way. We assume that the hardening timescale (the time
for two black holes to form hard binary) is rapid, that the loss
cone is full (Berczik et al. 2006, Sigurdsson 2003, Holley-Bockelmann
$\&$ Sigurdsson 2006), that no black holes are ejected, and that
there is no gas accretion involved. This gives us a strawman model
with which to compare more realistic black hole growth scenarios.
We assign seed black hole masses of 200$\Msun$. When two DMHs merge,
their seeds form a binary and merge at the center of the remnant
with the total combined mass of its progenitors. If more than two DMHs
merge, as often happens early on, the merger is sorted by DMH mass,
since dynamical friction is stronger on higher masses. Escala et al.
2006 suggests that the black hole binary will merge within a few
times 10$^7$ yr. Since DMHs typically merge in 10$^8$ yr, the MBHs
will coalesce soon after the DMHs merge in this model. As the halos
evolve to the present day, black holes can only grow in mass through
mergers, and in our (small cosmological volume) simulation reached a
maximum mass of M$_{\rm BH}$=3$\times10^5\Msun$. We find that this
value is too small to match observations.

For a sustained Eddington accretion, the mass growth rate is:

\begin{equation}
{\dot{M}_{\rm BH}}=\frac{M_{\rm BH}}{t_{\rm Sal}},
\end{equation}
where t$_{\rm Sal}$ is the Salpeter time-scale,
t$_{\rm Sal}$$\sim$4$\times10^7$ yr from the recent observations
(Hu et al. 2006). In our models, we assume that accretion is triggered by
galaxy mergers and lasts $\sim $t$_{\rm Sal}$. During this time, the black
holes approximately double their mass. However, we do not assume that every
merger supplies enough gas for the black hole to double its mass.
For example, during minor galaxy mergers (mass ratios much larger
than 10:1), the more massive galaxy will shred the satellite
(e.g. White 1983, Holley-Bockelmann $\&$ Richstone 1999), and the black
hole will sink toward the central black hole without a large
reservoir of gas. On the other hand, the merger of comparable
(mass ratios less than 10:1) galaxies will generate ample gas
accretion onto the central black hole, as the satellite triggers
a central starburst (Mihos $\&$ Hernquist 1994). Here we distinguish
two cases, depending on the mass ratio of merging DMHs. The first
is a more conservative criterion that allows black holes to accrete
gas if the mass ratio of host DMHs is less than 4:1 (4:1 accretion).
The second case sets an upper constraint on the mass of the final
black hole by allowing seeds to accrete gas as long as the merging
DMHs have a mass ratio less than 10:1 (10:1 accretion). In the
following section, we compare all three growth scenarios: black
hole growth through mergers only (dry growth); growth through
mergers combined with 4:1 gas accretion (4:1 growth); and finally,
growth through mergers combined with 10:1 gas accretion (10:1 growth).
These are simplifying assumptions made to explore parametrically the
range of solutions in our scenario.

\section{RESULTS}

From z=19 to z=12, we identified all of the DMHs in the manner
described in the previous section. With this approach, we
obtained the initial positions and formation redshifts for
1477 seed black holes, approximately 100 new MBHs per snapshot.
We assigned an initial mass of 200$\Msun$ to each MBHs and
traced their merger history from 0$\ltsim$z$\ltsim$19.

\subsection{Growth of Sagittarius A*}

Throughout the simulation, one DMH emerges as the largest in mass
and dominates the dynamics of its group. Its mass at z=0 is
M$_{\rm primary}$=4.5$\times$10$^{12}$$\Msun$ which roughly
corresponds to Andromeda's DMH, and is somewhat higher than estimates
for the Milky Way's halo (Dehnen et al 2006). This halo grows through
the smooth accretion of dark matter particles as well as DMH mergers.
Meanwhile, the central black hole grows through gas accretion
(if we included it) and mergers. We traced the growth of both the
primary halo and its massive black hole as a function of redshift
(Figure 1). Figure 1 compares the growth of both the black hole
(dry) and its host DMH. Overplotted is a descriptive redshift dependence
because it exists only where the curve has a negative slope. Initially,
the primary halo grows in mass only through smooth accretion and
reaches M$_{\rm primary}$=10$^9\Msun$ by z=15 (Fig. 1).
During this period, the black hole remains at its initial mass.
At all other redshifts, the growth of primary halo and any other DMH
can be described as a cycle of steady accretion of the surrounding dark
matter punctuated by rapid growth through mergers with incoming DMHs.

We present the DMH merger rates in Figure 2. In comparing Figure 1
and Figure 2, notice that changes in merger rate catalyze both the MBH
and primary halo mass growth. Figure 2 presents merger rates as the
number of mergers per unit time per unit redshift for all halos,
while Figure 1 traces the merger history for the primary halo only.
Nevertheless, since the primary halo participates in 90$\%$ of all
DMH mergers, Figure 2 essentially follows the evolution of the primary
halo. At z$\sim$11, the central black hole with mass
M$_{\rm BH}\sim3000\Msun$ has reached the IMBH range. At this
redshift, the merger rate reaches maximum (Fig 2.). Consequently,
the IMBH grows quickly from $\sim3000\Msun$ to $\sim10,000\Msun$
although surprisingly the bound DMH mass decreases.\footnote{In DMH mergers, 
the outskirts can become unbound for a short
period of time, but may be reaccreted as the structure
grows. The large number of DMH mergers between
10$\ltsim$z$\ltsim$12 (cf Figure 2), increases the number
of unbound dark matter particles, since there is not enough time for
reaccrete before another merger.} Merger rates decrease at z$\ltsim$10 but experience
two more peaks with periods of violent mergers around
z=6 and z=3 (Figure 2).

At z=0, there is one SMBH with mass
M$_{\rm SMBH}$=2.9$\times$10$^5$$\Msun$ at the center of the
primary halo (Figure 1 and Figure 3 thick line). This mass is too
small to match present day observations, so we know, as has been
stated, that mergers alone can not create a SMBH. As described
before, we used the Salpeter approximation to model different
gas accretion scenarios where the accretion efficiency depends
on the mass ratio of merging DMHs. In all three scenarios, the
black hole reaches the IMBH range at z$\sim$12 (Figure 3).
As the number of DMH mergers increases, the IMBH continues growing  --
faster, naturally, if gas accretion is more efficient. In a 4:1 growth scenario,
the IMBH grows to a $2.3\times10^6\Msun$ SMBH by z=5.5.
At these redshifts, low mass ratio mergers of DMHs are depleted and
SMBH growth through gas accretion stops. Therefore, the duty
cycle of the active galactic nucleus (AGN) hosting this SMBH dies at
z=5.5, too. In this model, the SMBH mass remains essentially constant
for $\ltsim$5.5, because it only experiences very high mass
ratio mergers with 200-1000$\Msun$ black holes. Finally, at z=0 the
AGN has evolved into galaxy hosting a $2.3\times10^6\Msun$ SMBH at
its center, strikingly similar to the mass of the Milky Way SMBH
(Schodel et al. 2002, Ghez et al. 2003, 2005). A highly efficient
10:1 growth scenario yields an upper constraint of $3.4\times10^7\Msun$
-- comparable to the SMBH observed in M31 (Bender et al. 2005).

Figure 4 shows the SMBH mass as a function of central velocity dispersion
and redshift for our three growth scenarios. We obtained the velocity dispersion
for dark matter at 10kpc comoving from the halo center. Recall that
10kpc comoving corresponds to 700pc at z=19 and 14 kpc at z=0. Similar
to Figure 3, every decrease in dark matter velocity dispersion is a
consequence of high merger rates at a specific redshift. The maximum
value for the velocity dispersion obtains at z=0, $\sigma$=120$\rm km^{-1}$.
When compared to the M-$\sigma$ relation from Gebhardt et al. 2001 and Merritt $\&$ Ferrarese 2001,
a SMBH of $\sim$3$\times10^7\Msun$ has a stellar velocity dispersion of
$\sigma$=120$\rm km^{-1}$ at z=0, which matches the SMBH mass in the 10:1
growth scenario. Of course our M-$\sigma$ relation is derived only
from the dark matter and at much larger radii, but this may give us a 
hint that the M-$\sigma$ relation is tied to the dynamics of the global 
potential. Exploring the formation of more massive SMBHs under this set 
of assumptions, including the redshift at which growth occurs and the 
merger rate as a function of redshift is beyond the scope of this paper.

\subsection{Black Hole Merger Rates and ULXs}

Although there are no studies of AGN duty cycle for z$>$6 and
M$_{\rm SMBH}$$\ltsim10^6\Msun$ (Wang et al. 2006), it is probable that
the AGN fueling mechanism -- gas accretion -- is similar to that of
higher mass galaxies.
The SMBH in 4:1 and 10:1 growth scenarios grows to its final mass
of M$_{\rm SMBH}$ $\sim 10^6\Msun$ - $few\times 10^7\Msun$ in the
redshift range 5.5$\ltsim$z$\ltsim$6. As major mergers dictate the
growth at z$>$6, the SMBH evolves as a low luminosity AGN
with the duty cycle governed by on and off switching of accretion
onto the black hole. The AGN duty cycle drops to zero at z$\sim$6
as the host galaxy becomes too large for major mergers to occur and
both the SMBH and its host galaxy enter a quiet phase of their growth
through mergers with much smaller galaxies.

Interestingly, all incoming small galaxies at z$\sim$6 in our model
carry an IMBH $\sim$1000$\Msun$.
This sets the stage for a phase where each new galaxy merger can be
characterized as a dwarf starburst galaxy with a central IMBH accreting
gas while sinking toward the SMBH at the center. In other words, in
these later phases the lower mass incoming black hole may accrete gas,
but we assume no significant gas accretion onto the central SMBH because
of the high mass ratio of the black holes. With the exception of the
largest black hole in our simulation, all other black holes fall into 
the IMBH range: 200$\Msun\ltsim$M$_{\rm IMBH}\ltsim$10$^4\Msun$. These black
holes are of the special interest as candidates for ULX sources
(Mii $\&$ Totani 2005). ULXs are interpreted as massive black holes
accreting gas in starburst galaxies, although this is still a matter
of debate (King et al. 2001).

Figure 5 shows the merger rates for
different mass ratio ranges and total binary masses. When
deconstructed into ranges of mass ratios and log total binary mass
(defined as p=log(m$_{\rm 1}$ + m$_{\rm 2}$)), these merger rates can be used to
predict the number of ULX sources in starburst galaxies throughout
the galaxy evolution, assuming that merging galaxies are reasonably gas rich.
For the mass ratio 1$\ltsim$m$_{\rm 1}$/m$_{\rm 2}$$<$10, we see
only minor differences between the three growth scenarios for
any range of binary mass (Fig 5a, 5b, 5c). For z$\gtsim$10
black holes grow mostly through mergers. For z$\sim$8, mass
ratio $<$10 and binary mass $10^3\Msun<$m$_{\rm BH}<10^4\Msun$
(3$\ltsim$p$<$4) the merger rate is R$\sim$10 yr$^{\rm -1}$ in all
three growth scenario -- and it reaches a maximum
R$_{\rm max}$$\sim$30 yr$^{\rm -1}$ for p$<$3. At lower redshifts
the merger rates decrease rapidly (Fig 5a, 5b, 5c), and gas accretion
becomes important (Figure 3). Nevertheless, there are no
changes in Fig 5a, 5b, 5c for z$<$10 since the mass ratio of merging
black holes becomes very large as the higher mass
black holes gain mass preferentially (Figure 6). In our scenario, only the 
black holes at the centers of merging DMHs grow through gas accretion; a 
large number of DMHs remain isolated for most of the simulation and merge 
late in high mass ratio encounters (middle and bottom panels of Fig 5). The
difference between the top and bottom panels in Figure 5 shows that
most of the black hole mergers at high redshifts are with
1$\ltsim$m$_{\rm 1}$/m$_{\rm 2}$$<$10 mass ratio.

Since black holes grow faster with more efficient gas accretion,
the merger rates for mass ratio 10$\ltsim$m$_{\rm 1}$/m$_{\rm 2}$$<$100
and total binary mass $10^4\Msun<$m$_{\rm BH}<10^6\Msun$, shift
toward higher redshifts in our accretion models (Fig 5d, 5e, 5f).
In this mass ratio range, 3$\ltsim$p$<$4 mergers peak at
8$\gtsim$z$\gtsim10$, with a merger rate of $\sim$10 depending on
the growth scenario. Very high mass ratio mergers
$10^2\ltsim$m$_{\rm 1}$/m$_{\rm 2}<10^4$ are
of the special interest (Fig 5. bottom panels). At z$<$6, high mass
ratio DMH mergers correspond to a dwarf galaxy or globular cluster
being consumed by massive galaxy. The IMBH carried by smaller
counterpart will eventually coalesce with SMBH at the center by
first forming a binary with total mass p$>$5 and high mass ratio
$10^2\ltsim$m$_{\rm 1}$/m$_{\rm 2}<10^4$. This is an ideal LISA source,
as will be explored in more detail in our next paper.

As a result, the merger rates in Figure 5d, 5e, 5f and Figure 7 for
the above total binary mass and mass ratio ranges directly correspond
to the number of IMBHs per galaxy per redshift. Since the accretion
timescale is approximately half the merger time scale,
the ULX number density is N$_{\rm ULX}$=R/2. In the case of dry growth
(Fig 5g), a black hole binary with mass $10^5\Msun<$m$_{\rm BH}<10^6\Msun$
has an approximately constant and high merger rate of R$\sim$15 yr$^{-1}$
for 2$<$z$<$6. This corresponds to $\sim$7 ULX
sources per starburst galaxy at 2$<$z$<$6.
Similarly the
$10^5\Msun<$m$_{\rm BH}<10^6\Msun$ binary in the 4:1 growth scenario has
R$\ltsim$10 yr$^{\rm -1}$ (Fig 5h) and N$_{\rm ULX}$=5.
More importantly Figure 5h shows N$_{\rm ULX}$$>$2 at 2$<$z$<$10.
For even higher mass ratios $10^4\ltsim$m$_{\rm 1}$/m$_{\rm 2}<10^5$, the
merger rate is R$\ltsim$12 yr$^{-1}$ for z$\ltsim$3.

Figure 7 shows merger rates for the 10:1 growth scenario and
high mass ratio $10^4\ltsim$m$_{\rm 1}$/m$_{\rm 2}<10^6$ for
$10^6\Msun<$m$_{\rm BH}<10^8\Msun$ mergers. ULX sources in Figure 7
and Figure 5g, 5h have constant number density of $\sim$10 per starburst
galaxy at z$>2$. The ULX number density drops rapidly at z=2. Note that
these ULX sources are not the same as the ULXB sources seen in local
universe with much shorter accretion lifetimes.

\subsection{Comparison to Results from Press-Schechter theory}

We compare our results with those obtained by various black hole
merger models from EPS theory. Figure 8 shows the merger rates
for four Press-Schechter models decribed in Sesana et al. 2007.
In the VHM model, massive DMHs (M$_{\rm DMH}$=10$^{11}$ - 10$^{15}$ $\Msun$)
are seeded with m$_{\rm BH}$$\sim$200$\Msun$ black holes at z=20; in the KBD model,
low mass halos (M$_{\rm DMH}$=10$^6$ - 10$^7$ $\Msun$) are seeded with 
m$_{\rm BH}$$\sim$5$\times10^4\Msun$ at 15$\ltsim$z$\ltsim$20; and the 
BVR models explore different redshift ranges for seeding black holes in
halos: m$_{\rm BH}$=$10^4 - 10^5 \Msun$
at 15$\ltsim$z$\ltsim$20 in the BVRhf model and 18$\ltsim$z$\ltsim$20
in the BVRlf model. Figure 8a shows superimposes their merger rates with our
results. Since our numerical simulations have smaller mass resolution than 
the VHM model, their DMHs (and therefore black holes) start merging later 
with a merger rate peaking at z=12. Our maximum merger rate is also higher 
because our range of seeding redshifts is larger 12$\ltsim$z$\ltsim$19. 
The mergers in question are small mass ratio mergers, with m$_{\rm BH}\ltsim10^4\Msun$
(Figure 8b). At z$\ltsim$10 our merger rates match the KBD model.
However, we have multiple peaks at z=6 and z=3 from episodes of
violent DMH dynamics that is not depicted by Press-Schechter theory
(Figure 8a). This is an advantage of numerical simulations, one that
acts to increase the merger rate, the black hole growth rate, and the 
mass ratio of typical mergers drastically.

In order to achieve $10^4\Msun\ltsim$m$_{\rm BH}\ltsim10^6\Msun$
(Figure 8c) with Press-Schechter models, one must start at higher
redshifts and assume higher initial black hole masses. The KBD model
starts with high mass seeds, and as a result the KBD merger rates
peak for $10^4\Msun\ltsim$m$_{\rm BH}\ltsim10^6\Msun$, and are larger
than the N-body rates, both in this mass range and at z$>$5. We ``spend''
BH seeds naturally during mergers so in this mass range our peak is
m$_{\rm BH}\ltsim10^4\Msun$ at z$>$5.

\section{DISCUSSION}

We used a high-resolution cosmological N-body simulation to study
the formation and growth of seed black holes into SMBHs and derived
black hole merger rates. We used physically-motivated formalisms
for seeding DMHs with black holes. Better understanding of the initial
mass function for Population III black holes will improve the accuracy
of our results in the future. Of course, black hole seeds do not
necessarily have to be Population III black holes. We can use our
approach to test different seeding scenarios. For example, it can be
used to set constraints for primordial black hole formation
(Mack et al. 2006).

Depending on the assumed growth scenario, we showed that gas accretion
combined with hierarchical mergers of massive black holes leads to
the formation of a Sagittarius A* type black hole in the center of
a Milky Way-sized DMH. In the case of very efficient gas accretion, an
M31-sized SMBH can form. In both cases, the SMBH reaches its final
mass at z$\sim$6. We showed that z=6 may be a critical redshift for
the transition from the AGN duty cycle dominated by high mass ratio
DMH mergers to a starburst galaxy phase where low mass ratio DMH
mergers supply a galaxy with a constant population of ULXs up to z=2. 
We argue that the z=6 turning point is consistent with the
maximum of AGN luminosity function at z=2 since the SMBHs in our
simulation have masses and luminosities below the currently observable
range. Also note that potential ULX sources in our simulation are at
z$>$2. They may correspond to nuclear clusters at the centers of
dwarf galaxies, and therefore likely sites of IMBHs formation. This
population is different from the low redshift ULX population which
may form from globular cluster dynamics and are likely undergoing
short lived binary accretion.

The final stages of the black hole merger are followed by an
emission of gravitational radiation, low in frequency but
relatively high in amplitude. In one of our follow up papers,
for each binary black hole inspiral and merger, the expected
gravitational wave signal for the LISA will be determined,
and the LISA event rate as a function of time calculated. In
particular, we will study whether LISA observations will be
able to distinguish between different assembly scenarios. One
interesting source we will be able to constrain will be
IMBH/SMBH mergers. We will calculate LISA detectability of
IMBH mergers for different mass ranges and binary mass ratios.
At this point, rough estimates can be made from Figure 4,
considering the known ranges for LISA sensitivity. In case of
dry growth (Fig 4g), a black hole binary with mass
$10^5\Msun<$m$_{\rm BH}<10^6\Msun$ is in the LISA range with
a very high merger rate of $\sim$15 yr$^{-1}$. Interestingly,
just as SMBHs dim electromagnetically at z=6, they turn on in
the gravitational wavebands.

There are a number of processes that might suppress massive black hole
merger rates. We assumed that every first star will produce black hole
as opposed to neutron star or pair detonation with no remnant. We also
assume that black holes merge efficiently and that recoil ejection
which is a function of spin, orientation, eccentricity, and mass ratio
of merging black holes is negligible. All of these processes will be
addressed in the follow up paper.

Finally, since our $O(10^{6})\Msun$ black hole was in place so early,
it might be tempting to interpret our results as an indictment against
downsizing, the recent observation that the number density of low mass
AGNs peaks at $z<1$ (Cowie et al. 2003, Merloni 2004, Heckman 
et al. 2004). However, we caution that
in order to achieve such high mass resolution, we simulated a small
volume, and are therefore plagued by small number statistics. In fact,
if we were to scale our simulation up in mass, so that the most massive
SMBHs were $O(10^{10})\Msun$, we find some indication that the more
isolated DMHs will host low mass SMBHs that gain most of their mass at late
redshifts compared to the most massive SMBH. We plan to explore the
growth of black holes as a function of environment and smooth accretion
efficiency in a follow up paper.

\section*{ACKNOWLEDGMENTS}

We would like to thank NASA's Columbia High End Computing program for
a generous time allocation, and the Center for Gravitational Wave
Physics at Pennsylvania State University for sponsoring this research.
This research was supported by a grant from the NSF, PHY 02-03046 and
from NASA's ATP NNG04GU99G. SS would also like to thank KITP at UCSB
and SLAC at Stanford University for their hospitality.

\clearpage

\begin{figure}
\vspace{0.5in}
\begin{center}
\includegraphics [width=3.in,angle=0]{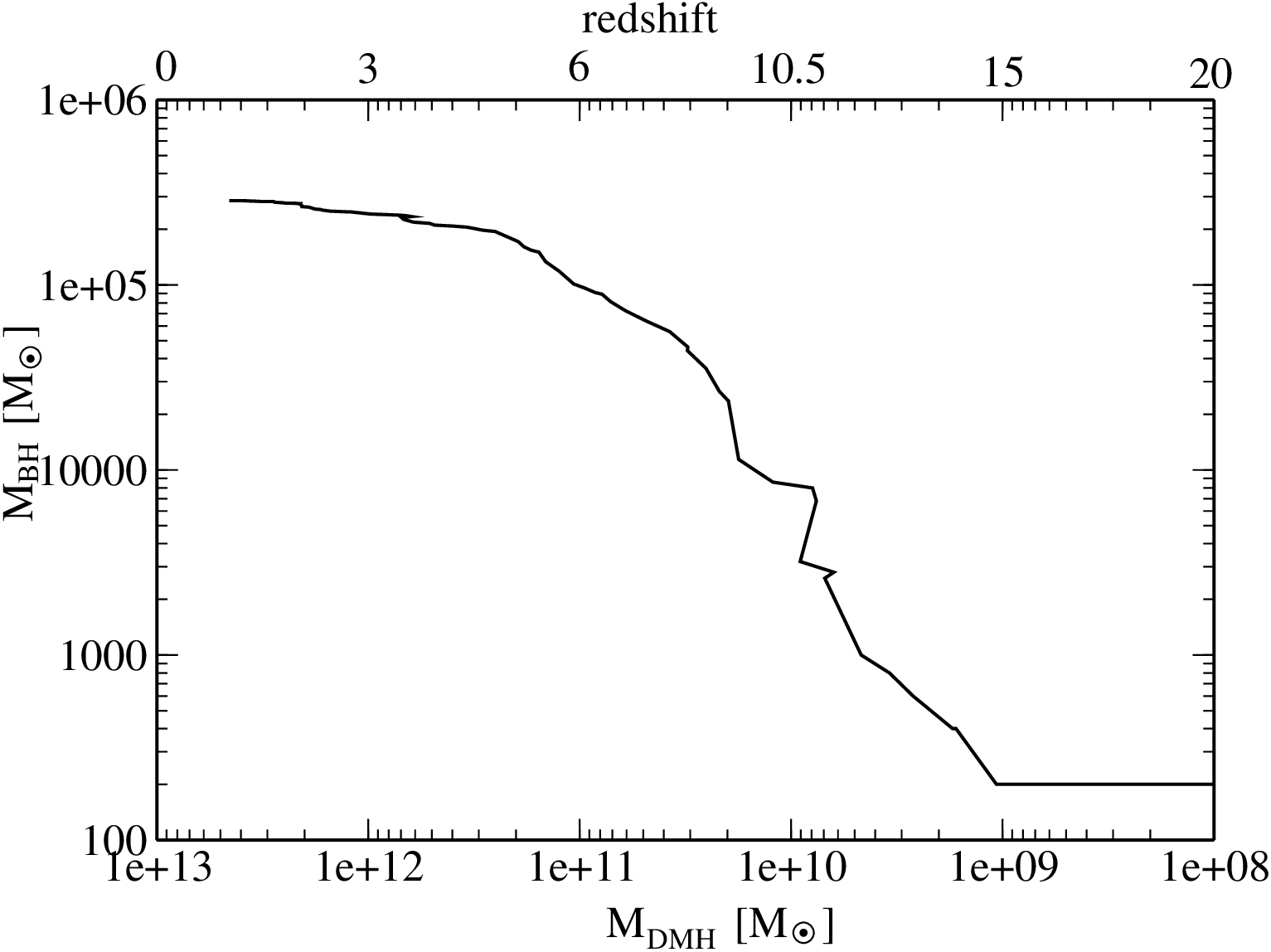}
\caption[Fig 1.]{Mass of the SMBH at the center of primary halo as a function
of primary halo's mass and redshift in the dry growth scenario. The growth of
primary halo and any other DMH can be described as a cycle of steady accretion
of the surrounding dark matter followed by rapid growth through mergers with
incoming DMHs. This process is best observed at z=11.}
\end{center}
\end{figure}

\begin{figure}
\vspace{0.5in}
\begin{center}
\includegraphics [width=3.in,angle=0]{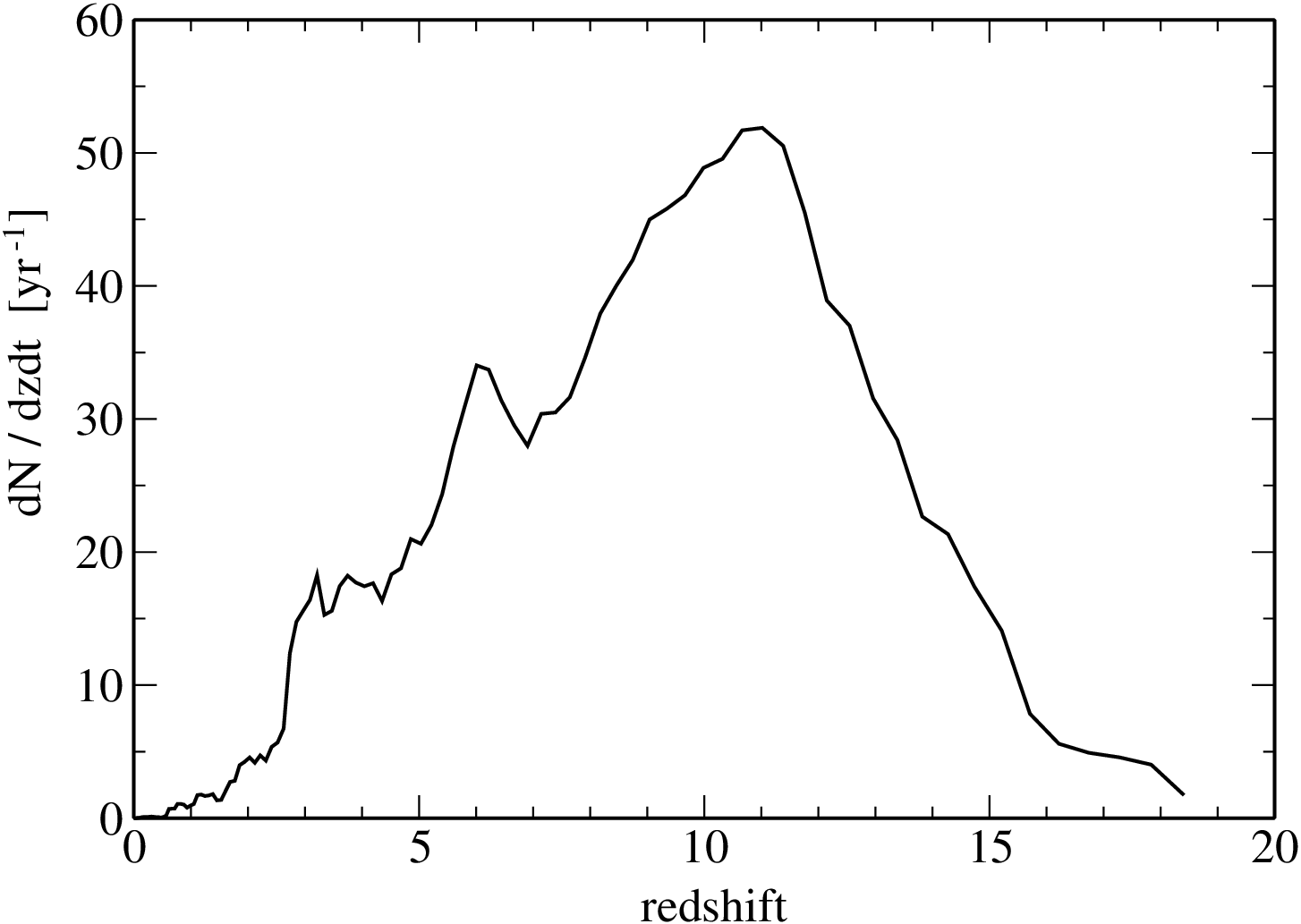}
\caption[Fig 2.]{Merger rates per unit time per unit redshift observed at z=0 as a
function of redshift. Merger rates reach maximum at z=11, decrease at redshifts
z$\ltsim$10 but experience two more peaks with periods of violent mergers around
redshifts z=6 and z=3.}
\end{center}
\end{figure}

\clearpage

\begin{figure}
\vspace{0.5in}
\begin{center}
\includegraphics [width=3.in,angle=0]{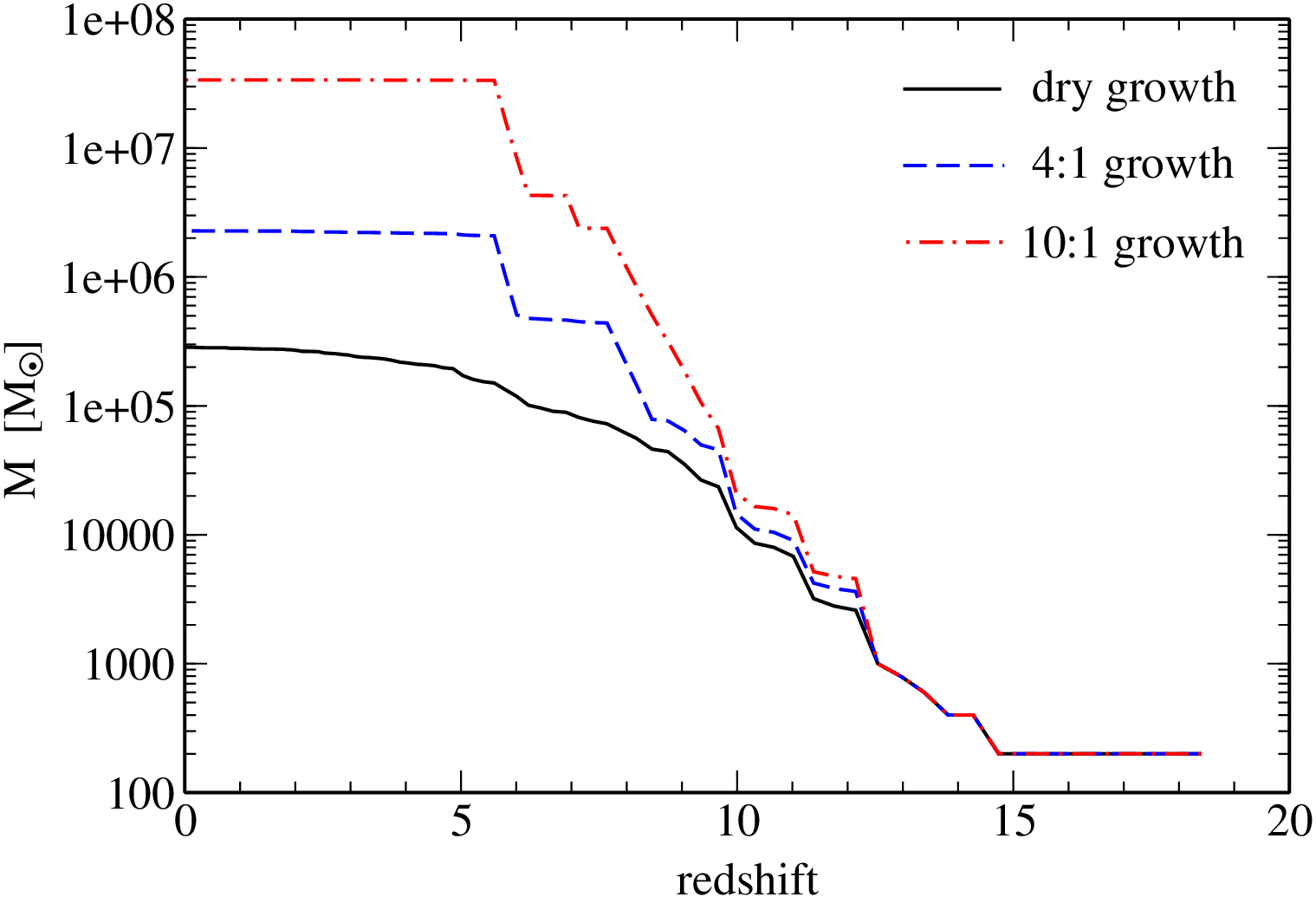}
\caption[Fig 3.]{Growth of SMBH in three different accretion scenarios presented
as mass of the SMBH as a function of redshift. Black hole mergers are dominant way
of growing IMBH at z$\gtsim$12. Gas accretion is important at 6$\ltsim$z$\ltsim$12.
At redshift z=6, SMBH reaches the maximum mass marking the transition of its host AGN
into the starburst galaxy.}
\end{center}
\end{figure}

\begin{figure}
\vspace{0.5in}
\begin{center}
\includegraphics [width=3.in,angle=0]{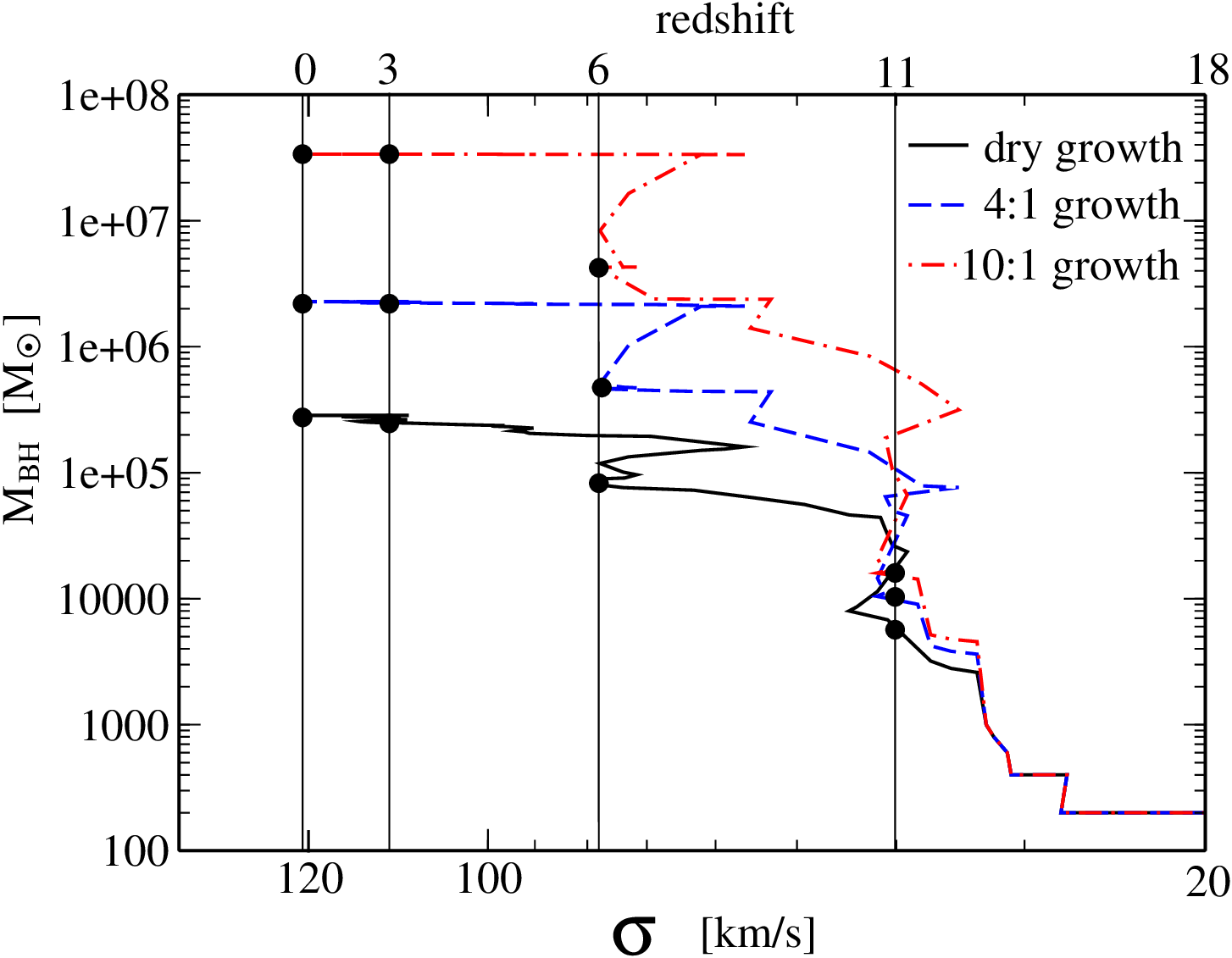}
\caption[Fig 4.]{Growth of SMBH in three different accretion scenarios presented
as mass of the SMBH as a function of dark matter central velocity dispersion at r=10kpc
($\sigma$), and as a function of redshift. Redshift dependence is descriptive hence
filled circles correspond to data points in terms of redshift. $\sigma$=120 $\rm km^{-1}$
at z=0 and it fits the M-$\sigma$ relation for 10:1 growth accretion.}
\end{center}
\end{figure}

\clearpage

\begin{figure}
\vspace{0.5in}
\begin{center}
\includegraphics [width=7.0in,angle=0]{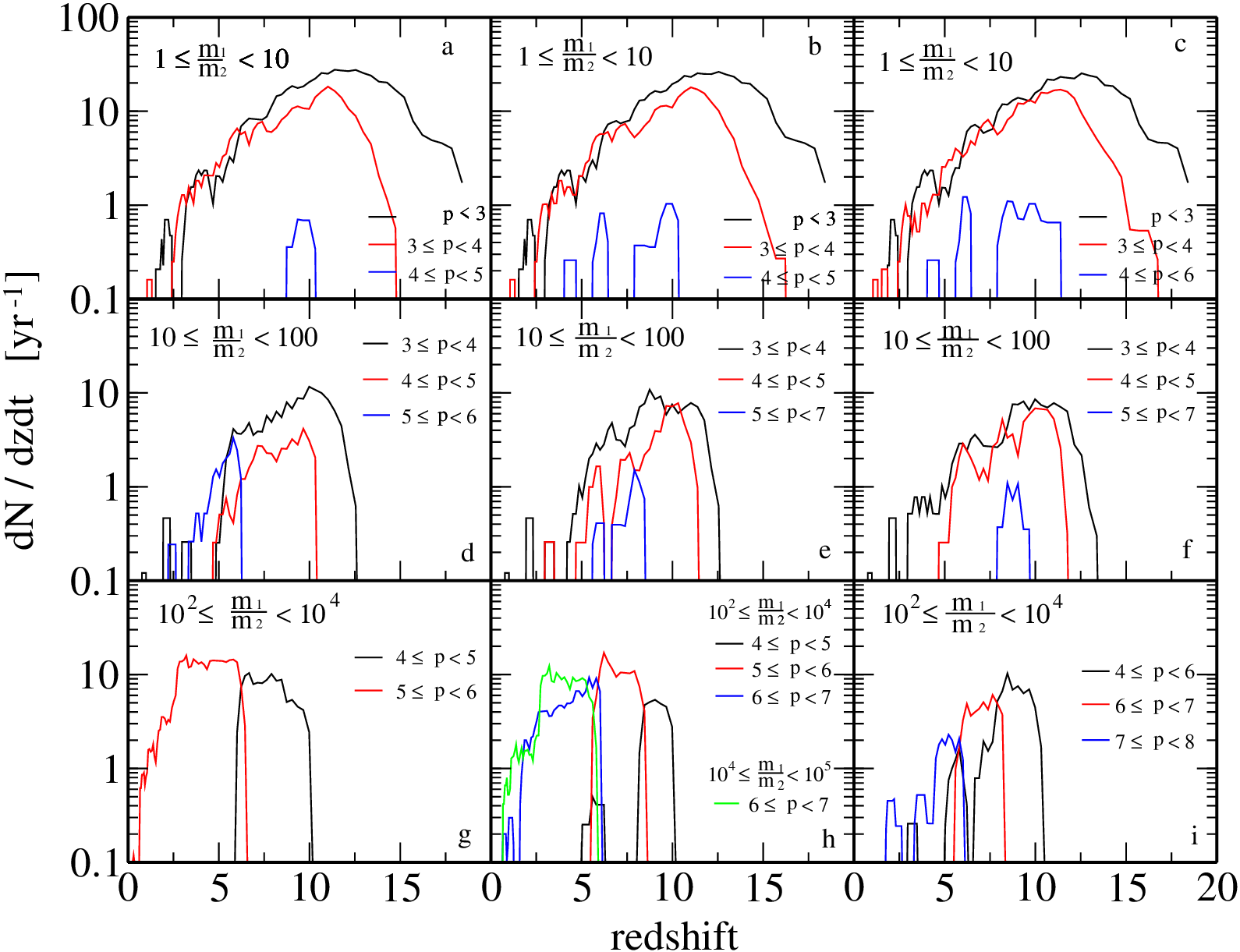}
\caption[Fig 5.]{IMBH merger rates as observed at z=0 as a function of redshift
and presented for different binary mass ratios and total binary mass ranges,
p = log (m$_{\rm 1}$ + m$_{\rm 2}$). 1$\ltsim$m$_{\rm 1}$/m$_{\rm 2}<10$ in a, b, c;
10$\ltsim$m$_{\rm 1}$/m$_{\rm 2}<100$ in d, e, f; 100$\ltsim$m$_{\rm 1}$/m$_{\rm 2}<10000$
in g, h, i; dry growth in a, d, g; 4:1 growth in b, e, h; 10:1 growth in c, f, i. 
Most of the mergers are low mass ratio mergers at z$>$10 . However, high 
mass ratio mergers at z$<$10 (g, h) are in the LISA range and have large merger 
rates R $\sim$ 15 yr$^{\rm -1}$ for a wide range of redshifts 2 $<$ z $<$ 10.}
\end{center}
\end{figure}

\clearpage

\begin{figure}
\vspace{0.5in}
\begin{center}
\includegraphics [width=3.0in,angle=0]{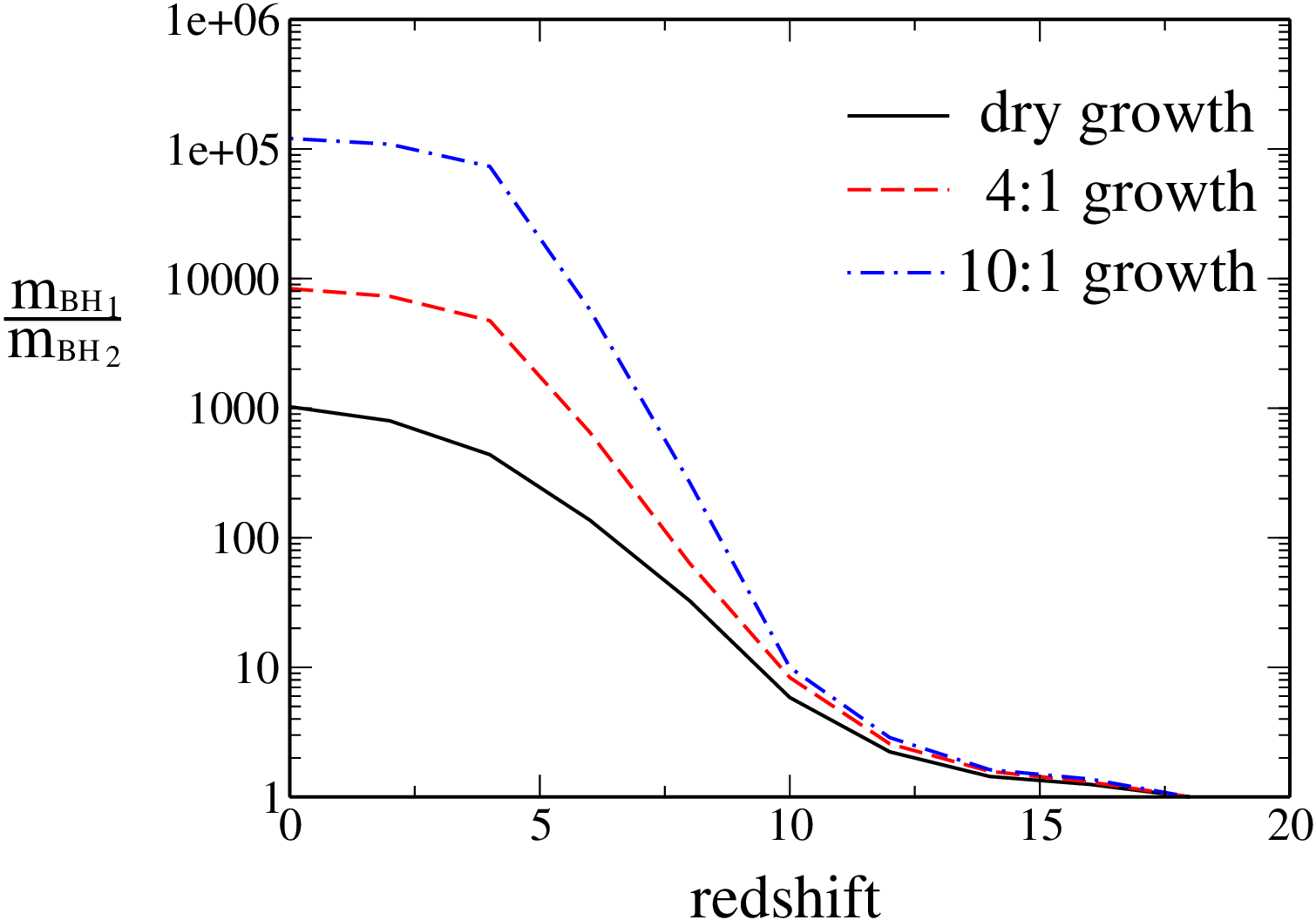}
\caption[Fig 6.]{Mass ratio of merging black holes averaged over all
mergers at a specific redshift for all three growth models. Black holes
grow mostly through mergers at z $>$ 10. Differences between three growing
models become apparent at z $<$ 10 where the increase in mass ratio of merging 
black holes is due to gas accretion.}
\end{center}
\end{figure}

\begin{figure}
\vspace{0.5in}
\begin{center}
\includegraphics [width=3.0in,angle=0]{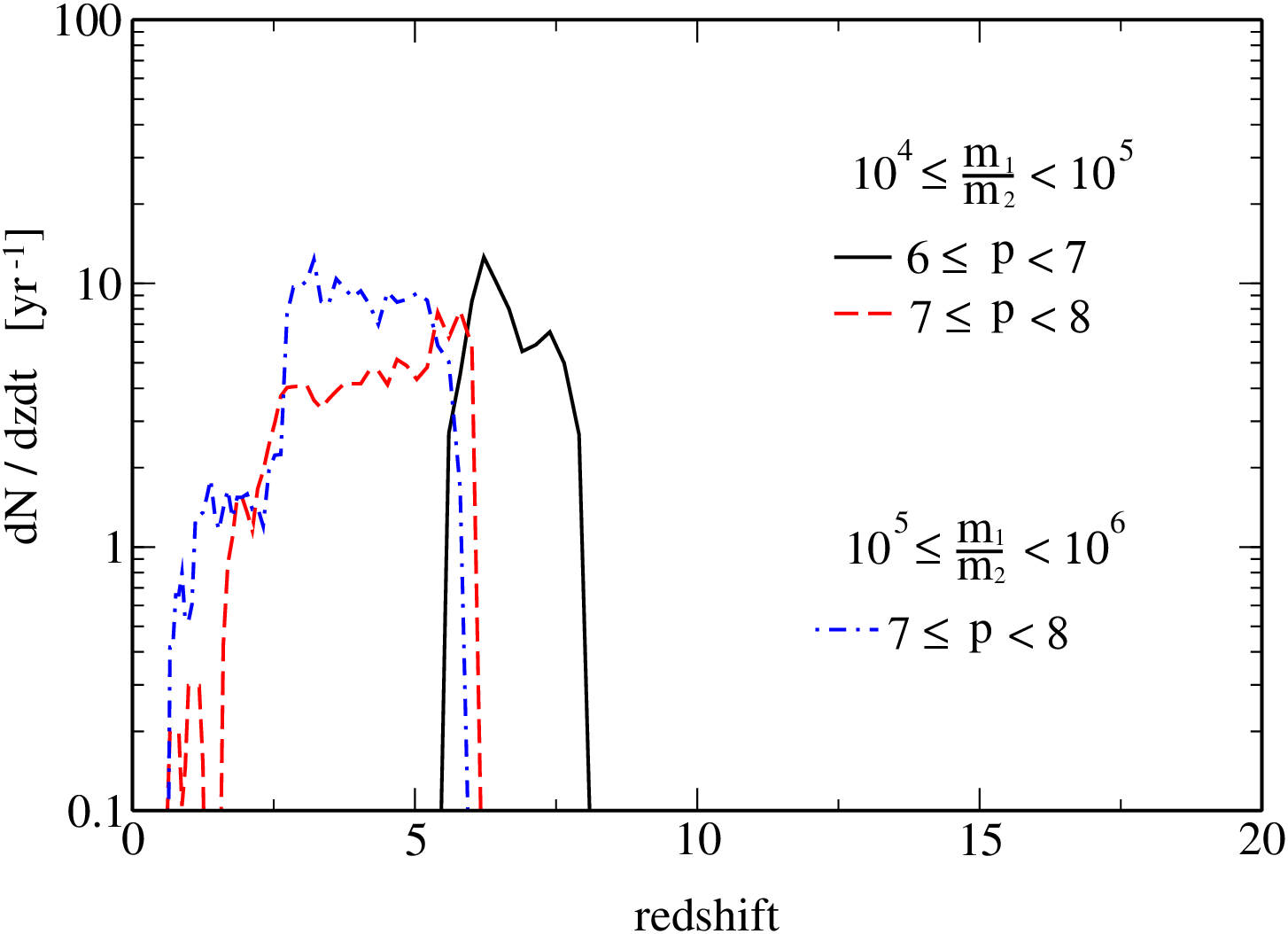}
\caption[Fig 7.]{Merger rates observed at z=0 as a function of redshift, presented is
the case of 10:1 growth scenario and merger rates for large binary mass and extreme
binary mass ratios, p = log (m$_{\rm 1}$ + m$_{\rm 2}$). Similar to Figure 5g and 5h,
mergers of SMBH with IMBH at z $>$ 2 will be observed by LISA, R $\sim$ 10 yr$^{\rm -1}$. }
\end{center}
\end{figure}

\clearpage

\begin{figure}
\vspace{0.5in}
\begin{center}
\includegraphics [width=7.0in,angle=0]{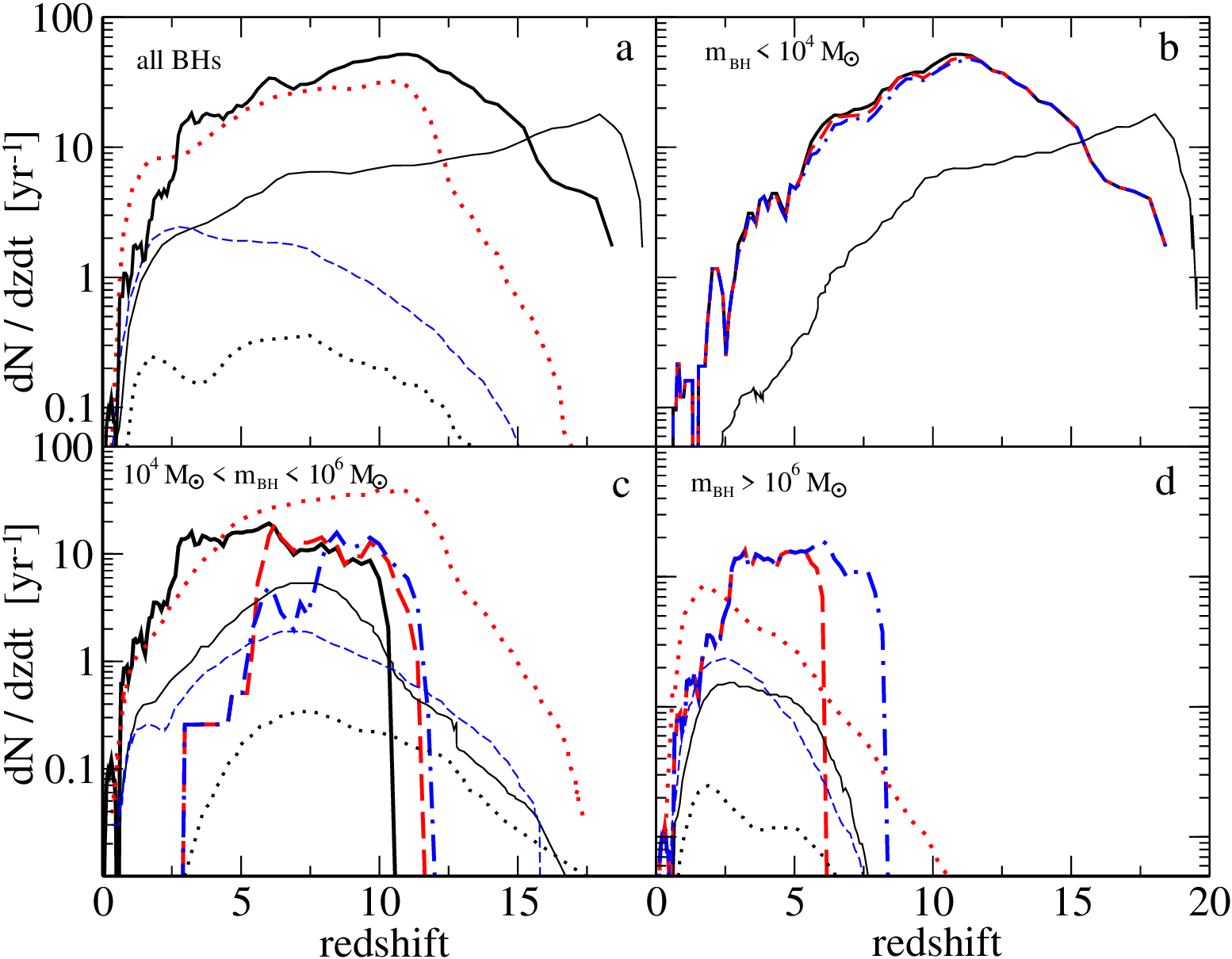}
\caption[Fig 8.]{Merger rates observed at z=0 as a function of redshift in different
binary mass intervals m$_{\rm BH}$=m$_{\rm 1}$ + m$_{\rm 2}$ for four
models described in Sesana et al. 2007, VHM in thin black line; KBD in dot-red line;
BVRlf in blue-dashed line and BVRhf in black-dot line. Overplotted are merger rates
from our N-body simulations for dry growth in thick black line; 4:1 growth in red
dashed line and 10:1 growth in blue dashed-dot line. We predict massive black hole 
merger rates to be larger than those predicted by semi-analitical models in the range
2 $<$ z $<$ 15. Numerical simulations also depict episodes of violent DMH dynamics 
which can be seen in multiple peaks at z=11; z=6; and z=3.}
\end{center}
\end{figure}


\begin{thebibliography}{dw}



\bibitem[Abel et al. 2000]{Abel2000}Abel, T., Bryan, G., \&
Norman, M., 2000, ApJ, 540, 39
\bibitem[Abel et al. 2002]{Abel2002}Abel, T., Bryan, G., \&
Norman, M., 2002, Sci, 295, 93A
\bibitem[Bender et al. 2005]{}Bender, R., Kormendy, J., Bower, G., 2005, ApJ, 631, 280
\bibitem[Berczik et al. 2006]{}Berczik, P., Merritt, D., Spurzem, R., Bischof, H.P., \&
2006, ApJ, 642L, 21B
\bibitem[Cowie et al. 2003]{}Cowie, L.L., Barger, A.J., Bautz, M.W., \& 
Brandt, W.N., Garmire, G.P., 2003, ApJ, 584, L5
\bibitem[Davis et al. 1985]{}Davis, M., Efstathiou, G., Frenk, C., White, S.D.M., \&
1985, ApJ, 292, 371
\bibitem[Dehnen et al. 2006]{Dehnen}Dehnen, W., McLaughlin, D.E. \& 
Sachania, J., 2006, MNRAS, 369, 1688
\bibitem[Erickcek et al. 2006]{}Erickcek, A.L., Kamionkowski, M., \&
Benson, A.J., 2006, MNRAS, tmp, 940E
\bibitem[Escala et al. 2005]{Escala}Escala, A, Larson, R.B., Coppi, P.S.,  \&
Mardones, D., 2005, ApJ, 630, 152E
\bibitem[Fabbiano 1989]{Fabbiano}Fabbiano, G., 1989, ARA\&A, 27, 87
\bibitem[Fabbiano $\&$ White 2006]{}Fabbiano, G., White, N.E., 2006,  \&
in Compact Stellar X-Ray Sources, ed. W. H. G. Lewin \& M. van der Klis \&
(Cambridge: Cambridge Univ. Press), 475
\bibitem[Gao]{}Gao, L., White, S.D.M., Jenkins, A., Frenk, C.S., Springel, V., \&
2005, MNRAS, 363, 379
\bibitem[Ghez et al. 2003]{}Gehz, A.M., Duchene, G., Matthews, K., 2003, ApJ, 586, 127
\bibitem[Ghez et al. 2005]{}Gehz, A.M., Salim, S., Hornstein, S.D., 2005, ApJ, 620, 744
\bibitem[Haehnelt]{}Haehnelt, M.G., 1994, MNRAS, 269, 199
Milosavljevic, M., ApJL, submitted, astro--ph/0306074
\bibitem[Heckman et al. 2004]{}Heckman, T.M., Kauffmann, G., Brinchmann, J., \&
Charlot, S., Tremonti, C., White, S.D.M., 2004, ApJ, 613, 109
\bibitem[Heger et al. 2003]{Heger}Heger et al., 2003, 2003, ApJ, 591, 288H
\bibitem[Holley-Bockelmann \& Sigurdsson 2006]{}Holley-Bockelmann, K., \&
Sigurdsson, S., 2006, astro-ph, 1520H
\bibitem[Holley-Bockelmann \& Richstone 1999]{}Holley-Bockelmann, K., \&
Richstone, D.O., 1999, ApJ, 517, 92H
\bibitem[Islam et al. 2003]{Islam}Islam, R. R., Taylor, J. E., Silk, J., \&
2003, MNRAS, 340, 647I
\bibitem[King et al. 2001]{King}King, A.R., Davies, M.B., Ward, M.J., Fabbiano, G. \& 
Elvis, M., 2001, ApJL, 552, 109
\bibitem[Kormendy \& Richstone 1995]{}Kormendy, J., Richstone, D., 1995, ARA\&A, 33, 581K
\bibitem[Lacey \& Cole]{}Lacey, C., Cole, S., 1993, MNRAS, 262, 627
\bibitem[Machacek et al. 2001]{}Machacek, M.E., Bryan, G.L., \&
Abel, T. 2001, ApJ, 548, 509
\bibitem[Mack et al.]{}Mack, K.J., Ostriker, J.P., Ricotti, M., astro-ph/0608642
\bibitem[Menou et al. 2001]{}Menou, K., Haiman, Z., Narayanan, V.K., \&
2001, ApJ, 558, 535
\bibitem[Merloni 2004]{}Merloni, A., 2004, MNRAS, 353, 1035
\bibitem[Merritt \& Ferrarese 2001]{}Merritt, D., Ferrarese, L., 2001, ApJ, 547, 140M
\bibitem[Mihos \& Hernquist 1994]{}Mihos, J.C., Hernquist, L., 1994, ApJ, 425L, 13M
\bibitem[Mii \& Totani]{}Mii, H., Totani, T., 2005, ApJ, 628, 873
\bibitem[Micic]{Micic}Micic, M., Abel, T., Sigurdsson, S., 2006, MNRAS, 372, 1540M
\bibitem[Nagashima et al. 2005]{}Nagashima, M., et al. 2005, ApJ, 634, 26N
\bibitem[Press \& Schechter]{PS}Press, W.H., Schechter P., 1974, ApJ, 187, 425
\bibitem[Ptak \& Colbert]{Ptak}Ptak, A., Colbert, E., 2004, ApJ, 606, 291
\bibitem[Reed et al. 2007]{}Reed, D.S., Bower, R., Frenk, C.S., Jenkins, A, Theuns, T., \&
2007, MNRAS, 374, 2R
\bibitem[Rhook \& Wyithe]{}Rhook, K.J., Wyithe, J.S.B., 2005, MNRAS, 361, 1145
\bibitem[Roberts \& Warwick]{Roberts}Roberts, T., Warwick, R., 2000, MNRAS, 315, 98
\bibitem[Sesana et al. 2004]{}Sesana, A., Haardt, F., Madau P., \&
Volonteri, M., 2004, ApJ, 611, 623
\bibitem[Sesana et al. 2007]{}Sesana, A., Volonteri, M., Haardt, F., \&
2007, astro-ph/0701556
\bibitem[Schneider et al. 2002]{Schneider}Schneider, R., Ferrara, A., \&
Natarajan, P., Omukai, K., 2002, ApJ, 571, 30
\bibitem[Schodel et al. 2002]{}Schodel, R., Ott, T., Genzel, R., 2002, Nature, 419, 694
\bibitem[Sigurdsson 2003]{}Sigurdsson, S., 2003, CQGra, 20S, 45S
\bibitem[Springel et al. 2001]{Springel}Springel, V., Yoshida, N., \&
White, S. D. M., 2001, NewA, 6, 79
\bibitem[Springel 2000]{PGF}Springel, V., 2000, MPA
\bibitem[Volonteri et al. 2003]{Volonteri}Volonteri, M., Haardt, F., \&
Madau, P., 2003, ApJ, 582, 559
\bibitem[Wang et al. 2006]{}Wang, J.M., Chen, Y.M., Zhang, F., 2006, ApJ, 647L, 17W
\bibitem[White 1983]{}White, S.D.M., 1983, ApJ, 274, 53W
\bibitem[Wise $\&$ Abel 2005]{Wise}Wise, J., H., Abel, T., 2005, ApJ, 629, 615W
\bibitem[Wyithe \& Loeb]{}Wyithe, J.S.B., Loeb, A., 2003, ApJ, 590, 691


\end{thebibliography}
\end{document}